\documentstyle[11pt]{article}
\input{psfig}
\newcommand{\mat}[1]{\mbox{\bf{#1}}}
\newcommand{\ket}[1]{\left | #1 \right \rangle}
\newcommand{\bra}[1]{\left \langle #1 \right |}
\newcommand{\amp}[2]{\left \langle #1 | #2 \right \rangle}
\newcommand{\proj}[1]{\ket{#1} \bra{#1}}
\newcommand{\tr}{\mbox{Tr} \,}
\begin{document}


\begin{center} {\large\bf Distinguishability of States and von Neumann Entropy }

Richard Jozsa$^\star$ and J\"{u}rgen Schlienz$^{\dagger
,}$\footnote{Present address: Rohde \& Schwarz GmbH \& Co. KG,
Test and Measurement Division, P.O. Box 80 14 69 D-81614 Munich,
M\"{u}hldorfstr. 15 D-81671 Munich, Federal Republic of
Germany.}\\[3mm] {\small $^\star$Department of Computer Science,
University of Bristol,\\ Merchant Venturers Building, Woodland
Road, Bristol BS8 1UB, UK.\\[2mm] $^\dagger$School of Mathematics
and Statistics, University of Plymouth,\\ Plymouth, Devon PL4 8AA,
UK.}
\end{center}


\begin{abstract}
Let $\{ \ket{\psi_1}, \ldots ,\ket{\psi_n}; p_1 ,\ldots ,p_n \}$
be an ensemble of pure quantum states. We show that it is possible
to increase all of the pairwise overlaps $|\langle \psi_i | \psi_j
\rangle |$ i.e. make each constituent pair of the states more
parallel (while keeping the prior probabilities the same), in such
a way that the von Neumann entropy $S$ is increased, and dually,
make all pairs more orthogonal while decreasing $S$. We show that
this phenomenon cannot occur for ensembles in two dimensions but
that it is a feature of almost all ensembles of three states in
three dimensions. It is known that the von Neumann entropy
characterises the classical and quantum information capacities of
the ensemble and we argue that information capacity, in turn, is a
manifestation of the distinguishability of the signal states.
Hence our result shows that the notion of distinguishability
within an ensemble is a global property that cannot be reduced to
considering distinguishability of each constituent pair of states.
\end{abstract}

Pacs Numbers: 03.67.-a

\section{Introduction}
The interpretation and physical significance of non-orthogonality
is one of the fundamental enigmas in the foundations of quantum
theory. Let $\ket{\phi}$ and $\ket{\chi}$ be two non-orthogonal
states of a quantum system. We may decompose $\ket{\phi}$ as a
superposition of components parallel and perpendicular to
$\ket{\chi}$ \begin{equation} \label{eq1} \ket{\phi} = a
\ket{\chi} + b\ket{\chi^\perp } \end{equation} where
$\amp{\chi^\perp}{ \chi} =0$ and $a=\amp{\chi}{ \phi}$. Since any
time evolution in quantum mechanics is unitary (when we include
also the state of any ambient environment) $\ket{\chi}$ and
$\ket{\chi^\perp}$ will evolve as though independent, remaining
orthogonal, and the decomposition in (\ref{eq1}) is preserved.
Thus we see that the overlap $|\amp{\chi }{ \phi}|$ measures the
extent to which the state $\ket{\phi}$ behaves as though it were
actually {\it equal} to the state $\ket{\chi}$. This view is
further formalised in the many worlds interpretation of quantum
theory according to which (\ref{eq1}) may be thought of as a
``splitting'' into two ``worlds''. In one of these worlds the
state $\ket{\phi}$ is indeed actually precisely {\it equal} to
$\ket{\chi}$.

The overlap (by which we will always mean the absolute value of
the inner product) is also a fundamental ingredient in the
question of (non-)distinguishability of quantum states. In
standard quantum measurement theory $|\amp{\chi}{\phi}|^2 $ is the
probability that $\ket{\phi}$ passes the test of ``being the state
$\ket{\chi}$''. Although $\ket{\phi}$ and $\ket{\chi}$ are {\it
distinct} states in the mathematical formalism of quantum theory,
there is no physical process that can distinguish them with
certainty  and indeed the overlap provides a quantitative measure
of the extent to which the states cannot be distinguished (as for
example, in the Peres measurement \cite{peres} for optimal
distinguishability of non-orthogonal states. This is in contrast
to distinct states in classical physics which are always perfectly
distinguishable in principle.

The purpose of this paper is to describe a situation which appears
to contradict the above intuitions. We will describe a situation
in which quantum states actually become more distinguishable (in a
certain natural sense) when they are made more parallel i.e. when
their overlap increases. Our notion of distinguishability will be
based on information--theoretic considerations and will rest on
the concept of von Neumann entropy. Recent work in quantum
information theory \cite{S95,JS,HJSWW,Barnum} has shown that this
alternative quantification of distinguishability is very natural
and compelling. Indeed if we view quantum states as carriers of
information then their capacity for embodying information is a
very natural measure of distinguishability i.e. a set of states
can communicate more information if and only if the states are
made more ``distinguishable'' (all else, such as prior
probabilities remaining the same). Quantum states may be used to
carry two different kinds of information, classical and quantum
information, and we will first briefly outline the essential
results characterising the respective information capacities which
form the basis of our quantification of distinguishability in
terms of von Neumann entropy.

Consider first the case of quantum information. Suppose that Alice
has a source which emits an unending sequence of qubit signal
states $\ket{\psi_1}= \ket{0}$ and
$\ket{\psi_2}=\frac{1}{\sqrt{2}} ( \ket{0}+\ket{1})$. Each
emission is chosen to be $\ket{\psi_1}$ or $\ket{\psi_2}$ with an
equal prior probability a half. Let $\rho =\frac{1}{2}
\proj{\psi_1} +\frac{1}{2} \proj{\psi_2}$ be the density matrix of
the source and let $S=S(\rho )= -\tr \rho \log_2   \rho $ be its
von Neumann entropy. Alice wishes to communicate the sequence of
states to Bob. Clearly this may be achieved by transmitting one
qubit per emitted state but according to the quantum source coding
theorem \cite{S95,JS,Barnum} she can communicate the quantum
information (with arbitrarily high fidelity) using
(asymptotically) only $S$ qubits per state. Furthermore this
compression is optimal: no fewer number of qubits per signal can
achieve this task. In our example a direct calculation gives that
$S= 0.601$ qubits per signal.

Now consider the analogous situation in classical physics: if we
have two classical signals with equal prior probabilities of a half
then no compression beyond 1 bit per signal is possible (by
Shannon's source coding theorem \cite{Cover}) and one may ask what
is the origin of the extra non-classical compression in the quantum
case? Clearly it is related to the overlap: if $\theta = \cos^{-1}
|\amp{\psi_1}{\psi_2}|$ is the angle between the two signal states
then $S$ depends only on $\theta$ and increases monotonically from
0 to 1 as $\theta$ increases from 0 to $\pi/2$ (corresponding to
the classical situation). In view of (\ref{eq1}) and the discussion
following it, one is tempted to think of $|\amp{\psi_1}{\psi_2}|$
as representing a redundancy or overlap of quantum information
between $\ket{\psi_1}$ and $\ket{\psi_2}$ which may be ``compressed
out'' i.e. to some extent $\ket{\psi_1}$ and $\ket{\psi_2}$ are the
``same'' and this common quantum information in every signal,
already known to Bob, need not be sent.

More generally if we have an ensemble of signal states $\{
\ket{\psi_1}, \ldots ,\ket{\psi_n} \}$ with prior probabilities
$p_1 , \ldots ,p_n $ then the quantum source coding theorem
asserts that the quantum information may be compressed to $S$
qubits per signal where $S$ is the von Neumann entropy of $\rho =
\sum_i p_i \proj{\psi_i}$ and that this compression is optimal.
Note that the von Neumann entropy $S(\rho )$ is always less than
or equal to the Shannon entropy $H(p_1 , \ldots ,p_n )$
\cite{HJW,Wehrl} and we might think of the extra quantum
compression to $S$ qubits beyond the classical limit of $H$ bits
per signal as being due to the overlap of the quantum information
represented by the constituent states as expressed in (\ref{eq1}).
Our results below will imply that this interpretation is
incorrect. Hence the origin of the extra quantum compression is
evidently more subtle.

Let $\{ \ket{\phi_1}, \ldots ,\ket{\phi_n}; p_1 , \ldots ,p_n \} $
denote the ensemble of quantum states $\ket{\phi_i}$ taken with
prior probabilities $p_i $ respectively. Let \[ \mbox{ $ {\cal E}
= \{ \ket{\psi_1}, \ldots ,\ket{\psi_n}; p_1 , \ldots ,p_n \}$ and
$\tilde{\cal E} = \{ \tilde{\ket{\psi_1}}, \ldots
,\tilde{\ket{\psi_n}}; p_1 , \ldots , p_n \} $} \] be two
ensembles with the same number of states  and with the same
corresponding prior probabilities. Let
\[ \rho = \sum_i p_i \proj{\psi_i} \hspace{5mm} \tilde{\rho}
= \sum_i p_i \tilde{\ket{\psi_i}} \tilde{\bra{\psi_i}} \] be the
respective density matrices and let $S$ and $\tilde{S}$ be the von
Neumann entropies. We will show that it is possible to have the
following situation: the states of $\cal E$ are all pairwise more
parallel (i.e. have greater overlap) than the corresponding states
of $\tilde{\cal E}$ yet the von Neumann entropy of $\cal E$ is
greater than that of $\tilde{\cal E}$ i.e. we simultaneously have
\[
\begin{array}{lcc} \mbox{(ENS1):} &
|\tilde{\bra{\psi_i}} \tilde{\psi}_j \rangle | \leq
|\amp{\psi_i}{\psi_j}| & \mbox{for all $i,j$}  \\
 \mbox{(ENS2):}  &   \tilde{S} < S  &  \end{array} \]
This is in contradiction to our intuitive discussion above. Each
{\em pair} of states of $\cal E$ has a greater overlap of quantum
information than the corresponding states of $\tilde{\cal E}$ in
the sense of (\ref{eq1}) yet as a {\it totality} they embody {\it
more} quantum information.

We will say that an ensemble ${\cal E}_2$ is a deformation of
${\cal E}_1$ if they have the same number of states and the two
ensembles have the same list of prior probabilities (in the given
order) i.e. the states of ${\cal E}_2$ are being thought of as
obtained by ``deforming'' the corresponding states of ${\cal E}_1$
while keeping the probabilities fixed.

We will show that (ENS1) with (ENS2) can never be satisfied for any
pair of ensembles ${\cal E}$ and $\tilde{\cal E}$ in {\it two}
dimensions (regardless of the number of states) but that for almost
all ensembles ${\cal E}$ of three states in three dimensions, there
is an ensemble $\tilde{\cal E}$ satisfying (ENS1) and (ENS2).

The phenomenon in (ENS1) and (ENS2) shows curiously, that
information capacity (or distinguishability) is a {\it global}
property of a set of states and not an accumulative local property
of pairs of constituent states. Indeed for ensembles $\{
\ket{\psi_1}, \ket{\psi_2}; p_1 ,p_2 \}$ of just two states, the
von Neumann entropy is a monotonically decreasing function of the
overlap $|\amp{\psi_1}{\psi_2} |$. Thus the overlap conditions in
(ENS1) imply that each constituent pair of signals has a {\it
diminished} capacity for information as we pass from $\tilde{\cal
E}$ to ${\cal E}$  yet (ENS2) states that the ensemble as a whole
develops an {\it increased} capacity.

Von Neumann entropy also characterises the {\it classical}
information capacity of an ensemble of quantum states. Suppose
that Alice is constrained to use the states $\ket{\psi_i}$ with
prior probabilities $p_i$ and she wishes to communicate classical
information to Bob. On receiving a string of states, Bob is
allowed to perform any joint measurement on a signal block of any
length in order to maximise his acquired mutual information about
the identity of the states. Then it may be shown \cite{HJSWW} that
the von Neumann entropy $S$ of the signal ensemble gives the
maximum amount of information per signal that Alice is able to
reliably transmit to Bob under the above constraints. Now {\it
classical} information capacity is very closely related to the
concept of distinguishability, which, by any definition, is itself
a form of classical information about the identity of the states.
Then (ENS1) with (ENS2) shows that the members of an ensemble of
quantum states can become pairwise less distinguishable yet as a
whole the ensemble becomes more distinguishable i.e. has more
classical information capacity.

\section{The Gram Matrix Formulation}
We now describe a formalism for studying the conditions
represented by (ENS1) and (ENS2) and give a method for generating
realisations in dimension $d>2$.

Consider an ensemble $\{ \ket{\psi_1}, \ldots ,\ket{\psi_n}; p_1 ,
\ldots ,p_n \}$ of $n$ states
 in $d$ dimensions. The density matrix is
\begin{equation} \label{rho} \rho = \sum_{i=1}^n
p_i \proj{\psi_i} \end{equation} We introduce the Gram matrix {\bf
G} defined as the $n\times n$ matrix of re-scaled inner products:
\begin{equation} \label{gr} G_{ij}= \sqrt{p_i p_j} \amp{\psi_i}{\psi_j}
\end{equation}
The Gram matrix enjoys the following two fundamental properties:
\begin{description}
\item[(G1):] The non-zero eigenvalues of {\bf G} are the same as the
non-zero eigenvalues of $\rho$ (and their respective
multiplicities are also the same). Note that in general $d\neq n$
so the mismatch in the numbers of eigenvalues is made up by zero
eigenvalues. It follows that $\rho$ and {\bf G} also have the same
von Neumann entropy.

To see this, introduce $n$ orthogonal vectors $\ket{e_i}$ in an
auxiliary Hilbert space and consider the pure state
\begin{equation} \label{rhog} \ket{\phi} = \sum_{i=1}^n \sqrt{p_i}
\ket{\psi_i} \ket{e_i} \end{equation} Then $\rho$ and {\bf G} are
just the two reduced states obtained by partial trace of
$\proj{\phi}$ over the second and first components respectively.
Hence they must have the same non-zero eigenvalues (c.f. appendix
of \cite{HJW}). $\Box$
\item[(G2):] {\bf G} is always a positive matrix and $\tr \mbox{\bf G}
=1$.  Conversely, if {\bf A} is any $m \times m$ positive matrix
with $\tr \mbox{\bf A}=1$ then {\bf A} is the Gram matrix of an
ensemble of $m$ states in $m$ dimensions \cite{HJ}.

The first part  follows immediately from eq. (\ref{rhog}) where
{\bf G} is identified as a density matrix itself. For the converse
statement note that if {\bf A} is positive we can write
$\mat{A}=\mat{B}^2$ where $\mat{B}$ is Hermitian so
$\mat{A}=\mat{B}^\dag \mat{B}$. Let $\hat{b_i}$ be the normalised
$i^{\rm th}$ column of $\mat{B}$ and let $t_i$ be its squared
length. Then $\mat{A}=\mat{B}^\dag \mat{B}$ expresses precisely
the fact that $\mat{A}$ is the Gram matrix of the ensemble of
$m$-dimensional states $\{ \hat{b_1},\ldots ,\hat{b_m};t_1 ,\ldots
,t_m \}$. The probabilities $t_i$ are just the diagonal entries of
{\bf A}.$\Box$
\end{description}
The Gram matrix, expressed in terms of the inner products rather
than the states themselves, provides a natural vehicle for studying
the conditions (ENS1) and (ENS2). Indeed we are generally not
interested in the actual positions of the ensemble states but only
in their relative positions i.e. inner products. The following
theorem shows that the Gram matrix encodes this information while
eliminating the superfluous data of overall unitary repositionings:
\\{\bf Lemma 1}: Two ensembles
\[ \mbox{ $ {\cal E}_1 = \{ \ket{\alpha_1}, \ldots ,\ket{\alpha_m}; p_1 ,
\ldots ,p_m \}$ on ${\cal H}_1$ and $ {\cal E}_2 = \{
\ket{\beta_1}, \ldots ,\ket{\beta_n}; q_1 , \ldots ,q_n \}$ on
${\cal H}_2$} \] have equal Gram matrices $G_1 = G_2$ if and only
if $m=n$, $p_i = q_i$ for $i=1, \ldots ,m$ and there is a unitary
transformation $U$ on ${\cal H}_1 \oplus {\cal H}_2$ with
$\ket{\beta_i} = U\ket{\alpha_i}$ for all $i$.$\Box$
\\We give the proof in the appendix.

 Let {\bf G} and $\tilde{\mat{G}}$ be respectively the Gram
matrices of the ensembles $\cal E$ and $\tilde{\cal E}$, which have
the same prior state probabilities (i.e. {\bf G} and
$\tilde{\mat{G}}$ have the same diagonals). Then (ENS1) is
equivalent to $|\tilde{G}_{ij} |
\leq |G_{ij}|$ for each $i$ and $j$, i.e.
\begin{equation} \label{r} \tilde{G}_{ij} = r_{ij} G_{ij} \hspace{3mm}
\mbox{ $r_{ij} \in {\cal C}$ and $|r_{ij}|\leq 1$ for each $i,j$}
\end{equation}
The matrix $\mat{r}= [r_{ij}]$ satisfies the following properties:
\begin{description}
\item[(R1)] $\mat{r}$ is Hermitian (without loss of generality)
\item[(R2)] the diagonal entries $r_{ii}$ are all equal to 1 and
$|r_{ij}| \leq 1$ for all $i$ and $j$.
\end{description}
However for given $\mat{G}$, $\mat{r}$ cannot be chosen
arbitrarily subject to (R1) and (R2) because $\tilde{\mat{G}}$ is
required to be a positive matrix (by (G2)). Also we wish to choose
$\mat{r}$ so that eq. (\ref{r}) induces a decrease in the entropy
of the Gram matrix $\mat{G}$.

The componentwise product of $\mat{G}$ and $\mat{r}$ in (\ref{r})
is known as the Hadamard product of matrices. We denote it by
\begin{equation} \label{hadam}
\tilde{\mat{G}} = \mat{r}*\mat{G} \end{equation}
 to distinguish it from the
usual matrix product. The Schur product theorem \cite{HJ} asserts
that the Hadamard product of any positive matrices is again a
positive matrix. Hence if $\mat{r}$ is chosen to be positive then
by (G2) $\tilde{\mat{G}}$ will again correspond to an ensemble of
states. However in this case (ENS2) can never be satisfied i.e.
the entropy is non-decreasing. To see this, let $\mat{G}$ be the
Gram matrix of an ensemble $\cal E$ comprising $n$ states
$\ket{\psi_i}$ with probabilities $p_i$. If $\mat{r}$ satisfying
(R1) and (R2) is {\it positive} then (G2) implies that
$\frac{1}{n}\mat{r}$ is also a Gram matrix of some collection of
states, $\ket{\xi_1},\ldots ,\ket{\xi_n}$ say, taken with equal
prior probabilities $\frac{1}{n}$. Thus \[ r_{ij}=
\amp{\xi_i}{\xi_j} \] Hence (\ref{r}) asserts that
$\tilde{\mat{G}}$ is the Gram matrix of the ensemble ${\cal E}(\xi
)$ comprising the states $\ket{\psi_i}\otimes \ket{\xi_i}$ with
probabilities $p_i$. Thus ${\cal E}(\xi )$ is an ``extension'' of
$\cal E$ obtained by simply adjoining the states $\ket{\xi_i}$ to
the corresponding $\ket{\psi_i}$'s. As such, the entropy $S(\xi )$
of ${\cal E}(\xi )$ can never be smaller than the entropy $S$ of
$\cal E$. We give three brief proofs of this fact, each invoking a
different (substantial) theorem. Firstly if $S(\xi ) <S$ then
Alice could reliably communicate the quantum information of $\cal
E$ to Bob using less than $S$ qubits per signal. She simply
adjoins the states $\ket{\xi_i}$, compresses to $S(\xi )$ qubits
per signal and on reception and decompression, Bob just discards
the extensions. This contradicts the quantum noiseless coding
theorem \cite{S95,JS,Barnum}. Secondly, in a similar way, $S(\xi )
<S$ contradicts the classical information capacity theorem
\cite{HJSWW}(which asserts that the von Neumann entropy is the
classical information capacity): the extended ensemble ${\cal
E}(\xi )$ cannot have a smaller information capacity since Alice
and Bob can always just ignore the presence of the extensions for
the purposes of classical communication. Thirdly, the passage from
$\ket{\psi_i}\ket{\xi_i}$ to $\ket{\psi_i}$ (i.e. discarding the
second state) is a physically realisable operation and hence a CP
map. Then Uhlmann's monotonicity theorem\cite{uhl77,lindblad}
(asserting that relative entropy can never increase under any CP
map) immediately implies that the entropy of $\cal E$ cannot
exceed the entropy of ${\cal E} (\xi )$ (since the entropy of any
ensemble of pure states $\rho_i = \proj{\phi_i}$ is just the
average relative entropy $\sum_i p_i S(\rho || \rho_i )$  where
$\rho =\sum_i p_i \rho_i$ is the ensemble density matrix).

Hence we have:\\{\bf Lemma 2:} If we wish to satisfy (ENS2)
together with (ENS1) it is necessary that the matrix $\mat{r}$ of
multipliers in eq. (\ref{r}) have at least one negative
eigenvalue.$\Box$

An example described later (having $\tilde{\mat{G}}$ positive)
will show that this necessary condition is unfortunately not also
sufficient.

\section{Ensembles in two dimensions}\label{sect3}

We begin by proving:\\{\bf Lemma 3:} (ENS1) and (ENS2) can never
be simultaneously satisfied for any ensembles in two
dimensions.$\Box$

For the case of two states ($n=2$) we have already noted that
lemma 3 follows readily from the explicit formula for von Neumann
entropy which, for any $p_1$ and $p_2$, depends monotonically on
the overlap of the two states. Alternatively in this case we note
that $\mat{r}$ is a 2 by 2 matrix which by (R1) and (R2) must take
the form
\[ \mat{r} = \left( \begin{array}{cc} 1 &  e^{i\beta}\cos \alpha \\
 e^{-i\beta}\cos \alpha & 1 \end{array} \right) \] This is always a positive
matrix (as the eigenvalues $1\pm \cos \alpha$ are both
non-negative) and we then apply lemma 2.

For general values of $n$ we introduce the linearised entropy
$S_{\rm lin}$ defined by \[ S_{\rm lin} = \tr (\rho - \rho^2 ) \]
Substituting (\ref{rho}) we get
\begin{equation} \label{SLIN} S_{\rm lin}=
(1-\sum_{i=1}^{n} p_i^2 ) -2 \sum_{i<j} p_i p_j
|\amp{\psi_i}{\psi_j}|^2 \end{equation} Hence $S_{\rm lin}$ is a
monotonically decreasing function of each of the overlaps
$|\amp{\psi_i}{\psi_j}|$. Next we note that for $d=2$, the von
Neumann entropy $S(\rho )$ is also a monotonically increasing
function of $S_{\rm lin}$, giving our claimed result that $S(\rho
)$ is a monotonically decreasing function of each overlap. To see
that $S$ is monotonically increasing with $S_{\rm lin}$ for $d=2$
let the eigenvalues of $\rho$ be $\lambda$ and $1-\lambda$. Then
\[ S= -\lambda \log \lambda - (1-\lambda ) \log (1-\lambda ) \]
and \[ S_{\rm lin} = \lambda + (1-\lambda ) - \lambda^2 -(1-\lambda
)^2 \] Computing $\frac{dS}{d\lambda}$ and $\frac{dS_{\rm
lin}}{d\lambda}$ shows that \[ \frac{dS}{dS_{\rm lin}} =
\frac{dS}{d\lambda}/\frac{dS_{\rm lin}}{d\lambda} >0. \]

\section{Ensembles in three dimensions}

We will focus on the case that the inequalities in (ENS1) are all
equalities: \begin{equation} \label{ens1eq} |\tilde{\bra{\psi_i}}
\tilde{\psi}_j \rangle | = |\amp{\psi_i}{\psi_j}| \hspace{3mm}
\mbox{for all $i,j$}
\end{equation} This will readily imply our basic result {\it viz}
that the entropy can be increased by increasing the overlap of each
pair of states. Indeed suppose that $\cal E$ and $\tilde{\cal E}$
are two different ensembles (with the same prior probabilities)
satisfying eq. (\ref{ens1eq}) but with $S \neq \tilde{S}$. Without
loss of generality suppose that $\tilde{S} <S$. Now consider a
small deformation of the states of $\cal E$ which slightly
increases all the pairwise overlaps, giving an ensemble ${\cal
E}^*$ with entropy $S^*$. Then either $S^* >S$ (so $S^{*} >
\tilde{S}$ giving a direct example of our result) or $S^* \leq S$.
In the latter case we will always have $S^*
>\tilde{S}$ if the deformation is sufficiently small, again giving
an example of our result (with ensembles $\tilde{\cal E}$ and
${\cal E}^*$).

We will show that for almost any ensemble $\cal E$ of three
linearly independent states (i.e. $n=3$ and $d=3$) it is possible
to deform $\cal E$, in two ways such that:
\begin{description}
\item[(D1)] all pairwise overlaps are increased and the entropy
increases,
\item[(D2)] all pairwise overlaps are decreased and the entropy
decreases.
\end{description}

For clarity in this section, it is useful to distinguish
(normalised) state  vectors, written as kets $\ket{\psi}$ from
physical states, which we will denote with square brackets as
$[\psi ]$. A physical state is a full set of all normalised vectors
that differ only in overall phase:
\[ [\psi ] = \{ e^{i\phi} \ket{\psi}: 0\leq \phi<2\pi \} \]
i.e. state vectors are elements of the Hilbert space whereas
physical states are elements of the projective Hilbert space. The
vectors $ e^{i\phi} \ket{\psi}$ in $[\psi ]$ are called phase
representatives of the state $[\psi ]$.

Consider three normalised vectors
$\ket{\psi_1},\ket{\psi_2},\ket{\psi_3}$ in ${\cal H}_3$. We have
the overlaps (non-negative real numbers):
\[ a_{12} = |\amp{\psi_1}{\psi_2}|
\hspace{5mm} a_{23} = |\amp{\psi_2}{\psi_3}|
\hspace{5mm} a_{31} = |\amp{\psi_3}{\psi_1}| \]
and the triple quantity denoted $\Upsilon_{123}$ (generally
complex) defined as
\begin{eqnarray} \Upsilon_{123}& = &
 \amp{\psi_1}{\psi_2}\amp{\psi_2}{\psi_3}\amp{\psi_3}{\psi_1} \nonumber \\
  & = & a_{12}a_{23}a_{31} e^{i\xi} \hspace{5mm} \mbox{
  for some phase $\xi$} \nonumber \end{eqnarray}
Note that the real numbers $a_{12}, a_{23},a_{31}, \xi$ are well
defined on physical states rather than just on the vectors i.e. if
we arbitrarily change the phases of the vectors then these four
numbers remain invariant. Also for any prior probabilities, the
density matrix and entropy of the ensemble $\{ \ket{\psi_1},
\ket{\psi_2}, \ket{\psi_3}; p_1 ,p_2 , p_3 \}$ is a function of
the corresponding physical states. Note also that unitary
transformations are well defined on physical states and leave the
quantities $a_{12}, a_{23},a_{31}, \xi$ invariant.

From the point of view of physics, we are primarily interested in
ensembles of physical states rather than ensembles of state
vectors. In contrast to the overlaps $a_{ij}$ and $\xi$, the Gram
matrix depends on the choices of phase representatives. For any
ensemble, given any choice of phase representative of $[\psi_1 ]$
we may always choose representatives of $[\psi_2 ]$ and $[\psi_3
]$ to make $\amp{\psi_1}{\psi_2}$ and $\amp{\psi_1}{\psi_3}$ real
and positive so the Gram matrix has the form: \begin{equation}
\label{gramform} \mat{G} = \left(
\begin{array}{ccc} p_1 & \sqrt{p_1 p_2 }a_{12} & \sqrt{p_1 p_3}
a_{31} \\ \sqrt{p_1 p_2} a_{12} & p_2 & \sqrt{p_2 p_3} a_{23}
e^{i\xi} \\ \sqrt{p_1 p_3} a_{31} & \sqrt{ p_2
p_3}a_{23}e^{-i\xi}& p_3
\end{array} \right) \end{equation}
To any ensemble of physical states we will associate a Gram matrix
of this form depending only on the invariants $a_{ij}$, $\xi$ and
the prior probabilities. Then if $\cal E$ and $\tilde{\cal E}$
have the same overlaps (as in eq. (\ref{ens1eq})), the Gram
matrices must be related as in eq. (\ref{hadam}) by a matrix of
multipliers $\mat{r}$ of the form
\begin{equation}
\label{rphi} \mat{r}(\phi )= \left(
\begin{array}{ccc} 1 & 1 & 1 \\ 1 & 1 & e^{i\phi} \\ 1 &
e^{-i\phi} & 1 \end{array} \right)
\end{equation}

To study deformations preserving overlaps we begin by giving a
characterisation of the set of all possible triples of physical
states $\{ [\psi_1 ], [\psi_2 ], [\psi_3 ] \}$ compatible with a
given prescribed set $a_{12},a_{23},a_{31}$ of overlaps. We
clearly have any overall unitary transformation of any allowed
triple but we are especially interested in triples that are not
unitarily related. A complete characterisation is given in the
following theorem, whose proof is given in the appendix.
\\{\bf Theorem 1:} Suppose we are given real numbers
\[ 0\leq a_{12} \leq 1 \hspace{5mm} 0\leq a_{23} \leq 1 \hspace{5mm}
0\leq a_{31} \leq 1 \]
\begin{description}
\item[(a)] If $\{ [\psi_1 ], [\psi_2 ], [\psi_3 ] \}$ is any set of
physical states having $a_{12}, a_{23}, a_{31}$ as overlaps then
the phase $\xi$ of $\Upsilon_{123}$ satisfies
\begin{equation} \label{xi}
1+2a_{12}a_{23}a_{31}\cos \xi \geq a_{12}^2 + a_{23}^2 + a_{31}^2
\end{equation}
Conversely, for any solution $\xi$ of eq. (\ref{xi}) there is a
set of states having overlaps $a_{12}, a_{23}, a_{31}$ and triple
quantity $\Upsilon_{123} = a_{12}a_{23}a_{31} e^{i\xi}$. Thus the
existence of a solution of eq. (\ref{xi}) is a necessary and
sufficient condition for a set of real numbers $a_{12},
a_{23},a_{31}$ to be realisable as a set of overlaps.
\item[(b)] The solutions $\xi$ of eq. (\ref{xi}) give a
one-to-one parameterisation of all sets of states up to unitary
equivalence, that have the prescribed overlaps. We always take
$\xi$ to be in the interval $[-\pi,\pi ]$ so if eq. (\ref{xi}) has
solutions they are always of the form $-\xi_{max} \leq \xi \leq
\xi_{max}$ with $\xi_{max} \leq \pi$ (and we identify the values
$\pm \pi$ if $\xi_{max} = \pi$).
\item[(c)] (Explicit formulae up to unitary equivalence). Suppose that
$\{ [\psi_1 ], [\psi_2 ], [\psi_3 ] \}$ is any set of physical
states having $a_{12}, a_{23}, a_{31}$ as overlaps. Then there is
an orthonormal basis $\{ \ket{0}, \ket{1}, \ket{2} \}$ of ${\cal
H}_3$ such that phase representatives of the states are given by:
\begin{equation} \label{xi123} \begin{array}{l}
\ket{\psi_1 (\xi)} = \ket{0}
\\
\ket{\psi_2 (\xi)} = a_{12} \ket{0} + \sqrt{1-a_{12}^2} \ket{1}
\\
\ket{\psi_3 (\xi)} =  a_{13}\ket{0} +
\frac{a_{23}e^{i\xi} - a_{23}a_{31}}{\sqrt{1-a_{12}^2}} \ket{1} +
\frac{\sqrt{1-a_{12}^2 -a_{23}^2 - a_{31}^2 +
2a_{12}a_{23}a_{31}\cos \xi }}{\sqrt{1-a_{12}^2}} \ket{2}
\end{array}
\end{equation}
Here $\xi$ is the phase of $\Upsilon_{123}$.$\Box$
 \end{description}
 To illustrate theorem 1 and its significance for further developments
 we describe an example. \\{\bf Example 1.} Consider the ensemble ${\cal E}
= \{ \ket{\alpha_1}, \ket{\alpha_2}, \ket{\alpha_3}; \frac{1}{3},
\frac{1}{3},\frac{1}{3} \}$ where
\begin{equation} \begin{array}{ccl}
\ket{\alpha_1} &= & \ket{0} \\
\ket{\alpha_2} &= & \frac{1}{\sqrt{2}} \left( \ket{0}+\ket{1} \right) \\
\ket{\alpha_3} & = & \frac{1}{\sqrt{3}} \left( \ket{0}+\ket{1}
+\ket{2} \right)
\end{array} \end{equation}
The Gram matrix has eigenvalues 0.053, 0.145 and 0.802 with entropy
$S= 0.613$ (where natural logarithms have been used). According to
theorem 1, up to unitary equivalence any ensemble with the same
overlaps has the form ${\cal E}(\xi) = \{ \ket{\alpha_1 (\xi)},
\ket{\alpha_2 (\xi )}, \ket{\alpha_3 (\xi)};\frac{1}{3},
\frac{1}{3},\frac{1}{3} \}$ where
\begin{equation} \label{alpxi}\begin{array}{ccl}
\ket{\alpha_1 (\xi)} &= & \ket{0} \\
\ket{\alpha_2 (\xi)} &= & \frac{1}{\sqrt{2}} \left( \ket{0}+\ket{1} \right) \\
\ket{\alpha_3 (\xi)} & = & \frac{1}{\sqrt{3}} \ket{0}+
\frac{2e^{i\xi}-1}{\sqrt{3}}\ket{1}
+ \sqrt{\frac{4}{3}\cos \xi -1}\ket{2}
\end{array} \end{equation}
Here the parameter $\xi$ is constrained by eq. (\ref{xi}) giving:
\[ \cos \xi \geq \frac{3}{4} \hspace{5mm} \mbox{i.e. $\xi_{max} = 0.72$ radians}
\]
$\mat{G}(\xi)$, the Gram matrix of ${\cal E}(\xi )$, is positive
so long as $|\xi | \leq \arccos \frac{3}{4}$. At the maximum value
of $\xi$ the amplitude of $\ket{2}$ becomes zero and the states
become linearly dependent.

The von Neumann entropy of ${\cal E}(\xi)$ may be computed from the
eigenvalues of $\mat{G}(\xi)$. This is shown in figure 1 and we see
that the entropy falls monotonically with $\xi$ for the ensembles
${\cal E}(\xi)$ which have constant overlaps. (For negative values
of $\xi$ the graph is reflected in the vertical axis). Thus for any
$0<\xi < \arccos \frac{3}{4}$ we may deform ${\cal E}(\xi)$ to
obtain an ensemble ${\cal E}^*$ with the same overlaps but having
strictly lower (respectively higher) entropy. Then by slightly
deforming ${\cal E}^*$ to make all overlaps lower (respectively
higher) we obtain a deformation of ${\cal E}(\xi)$ satisfying {\bf
(D2)} (respectively {\bf (D1)}). Note that  if $\xi
= 0$ (i.e. $\Upsilon_{123} $ is real) then we only get {\bf (D2)} ( and not
{\bf (D1)}) by {\em this} method. Also for $\xi = \xi_{max}=\arccos
\frac{3}{4}$ we get only {\bf (D1)} (by {\em this} method)
as $\xi$ can only be decreased.$\Box$

\begin{figure}
 \centerline{
   \psfig{file=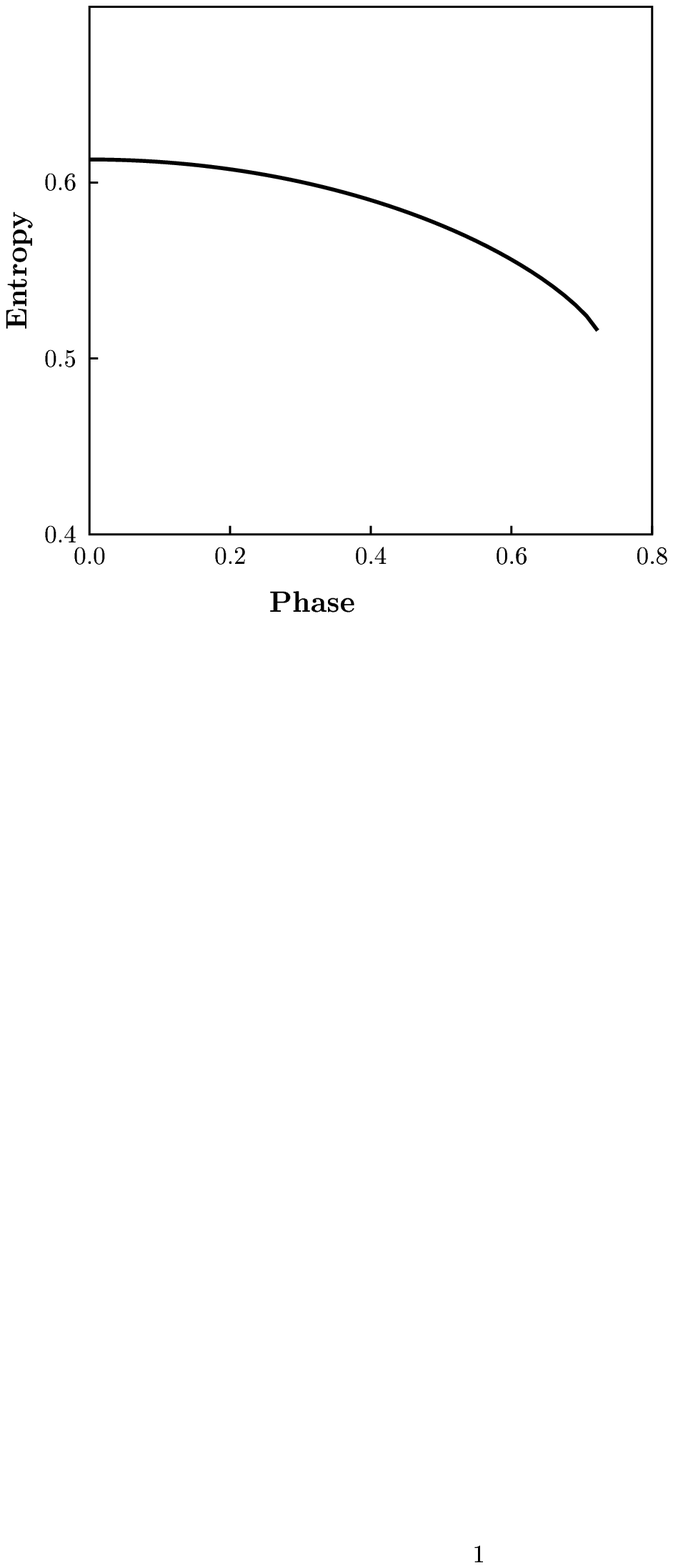,width=0.7\textwidth}}
  \caption{Variation of von Neumann entropy for the family of ensembles in
  eq. (\ref{alpxi}) with unchanging overlaps, as a function of $\xi$, the
  phase of $\Upsilon_{123}$. The maximum value of $\xi$ is $\arccos
  \frac{3}{4} \approx 0.72$ radians.}
\end{figure}

We now return to general ensembles of three states and study the
variation of von Neumann entropy with deformations that preserve
the overlaps. According to Theorem 1 these deformations, up to
unitary equivalence, are parameterised by $\xi$. We will find that
the behaviour exhibited in example 1 is generic -- the entropy
$S(\xi )$ falls monotonically as $\xi$ increases from 0 to
$\xi_{max}$ just as in figure 1. Then the same deformation
arguments as in example 1 give the following theorem:
\\{\bf Theorem 2:}
 Suppose the ensemble ${\cal E}=
 \{ \ket{\psi_1}
,\ket{\psi_2}, \ket{\psi_3}; p_1 ,p_2 ,p_3 \}$ has rank 3 (i.e.
the un-normalised states $\sqrt{p_i} \ket{\psi_i}$ are linearly
independent) with overlaps $a_{12},a_{23},a_{31}$ and phase
$\xi_0$ of $\Upsilon_{123}$. \begin{description}
\item[(a)] If $\Upsilon_{123}$ is not real ( i.e. $\xi_0 \neq 0,\pi$) then
$\cal E$ can be deformed according to both {\bf (D1)} and {\bf
(D2)}.
\item[(b)] If $\Upsilon_{123}$ is real positive (i.e. $\xi_0 =0$) then
$\cal E$ can be deformed according to {\bf (D2)}. If
$\Upsilon_{123}$ is real negative (i.e. $\xi_0 =\pi$) then $\cal
E$ can be deformed according to {\bf (D1)}.$\Box$
\end{description}
The proof of theorem 2 is given in the appendix.

Note that in theorem 2(b) we have not ruled out the possibility of
a {\bf (D1)} deformation (respectively {\bf (D2)} deformation)
when $\cal E$ has $\Upsilon_{123}$ real positive (respectively
negative). We have shown only that such deformations cannot be
achieved by the particular method of first altering the entropy
while keeping the overlaps constant and then slightly increasing
or decreasing the overlaps. Indeed consider the ensemble $\cal F$
of states:
\[ \begin{array}{l} \ket{\psi_1} = \ket{0} \\
\ket{\psi_2}= \frac{1}{\sqrt{2}}( \ket{0} + \ket{1}) \\
\ket{\psi_3} = \frac{\sqrt{2}-1}{6}\ket{0} + \frac{\sqrt{2}+1}{6} \ket{1}
\end{array} \]
taken with equal probabilities. Here we have $\Upsilon_{123}$ real
positive so our method cannot give a {\bf (D1)} deformation.
However such deformations do exist, for example the ensemble ${\cal
F}'$ of states:
\[ \begin{array}{l} \ket{\psi_1} = \ket{0} \\
\ket{\psi_2}= \frac{1}{\sqrt{2}}( \ket{0} + \ket{1}) \\
\ket{\psi_3} = \frac{2}{3\sqrt{2}}\ket{0} + \frac{4}{3\sqrt{3}} \ket{1}
+ \frac{\sqrt{7}}{3\sqrt{3}}\ket{2}
\end{array} \]
${\cal F}'$ has overlaps greater than or equal to those of $\cal F$
yet its entropy $S' = 0.91$ is greater than the entropy $S=0.85$ of
$\cal F$ i.e. ${\cal F}'$ is a deformation of $\cal F$ of type {\bf
(D1)}.

This leads us to conjecture that the conclusions in theorem 2(a)
also hold for the ensembles in (b) and more generally that any
ensemble containing a subset of three states that has no parallel
or orthogonal pairs, admits deformations of both types {\bf (D1)}
and {\bf (D2)}. However it is possible to show from our results
that if such deformations always exist, they are {\em not} always
realisable as {\em continuous} deformations of the appropriate
type starting with the given ensemble. Indeed let $\cal E$ be any
ensemble with $\Upsilon_{123}$ complex, consisting of three
linearly dependent states, lying in a plane $P$ (so that from
theorem 1(c) the phase $\xi$ of $\Upsilon_{123}$ must take the
maximum value allowed by the overlaps). From the result of section
\ref{sect3} we know that there are no deformations (continuous or
not) of type {\bf (D1)} or {\bf (D2)} lying within the plane $P$.
Since $\cal E$ is planar, its density matrix has a zero
eigenvalue, say $\lambda_3 = 0$. The entropy function $S(\lambda_1
, \lambda_2 , \lambda_3 )$ has infinite slope at $\lambda_3 =0$ in
the $\lambda_3$ direction (and finite slopes in the $\lambda_1$
and $\lambda_2$ directions, for $\lambda_1 , \lambda_2
>0$) so any continuous deformation of $\cal E$ out of the plane $P$
must begin to {\em increase} $S$, regardless of the overlaps. Hence
no {\em continuous} deformations, by any method whatever, of type
{\bf (D2)} can exist (yet there may still exist a ``distant''
ensemble ${\cal E}'$ with decreased overlaps and decreased entropy
(i.e. a {\bf (D2)} deformation) which cannot be connected to $£\cal
E$ by a continuous family of {\bf (D2)} deformations.

In relation to theorem 1 it is interesting to note that in
\cite{MP} Gisin and Popescu have described another method for
(discontinuous) deformation of a particular class of ensembles
that preserves all overlaps. Let $\ket{\underline{n}}$ denote the
qubit state corresponding to the unit vector $\underline{n}$ on
the Bloch sphere. Let ${\cal E} = \{ \ket{\underline{n}_1}
\ket{\underline{n}_1} , \ldots ,\ket{\underline{n}_k}
\ket{\underline{n}_k}; p_1 \ldots ,p_k \}$ be any ensemble of two
qubit states where each constituent state comprises two copies of
a single qubit state. We now replace each second component
$\ket{\underline{n}_i}$ by the orthogonal state
$\ket{-\underline{n}_i}$ to get ${\cal E}^*  = \{
\ket{\underline{n}_1} \ket{-\underline{n}_1} , \ldots
,\ket{\underline{n}_k} \ket{-\underline{n}_k}; p_1 \ldots ,p_k
\}$. Clearly $\cal E$ and ${\cal E}^*$ have the same pairwise
overlaps but in general the von Neumann entropies will be
different. Indeed $\cal E$ always lies within the three
dimensional subspace of symmetric states whereas ${\cal E}^*$
generally spans the full four dimensions of two qubits. In
\cite{MP} the ensemble $\cal E$ uniformly distributed over the
whole Bloch sphere was considered. It was shown that ${\cal E}^*$
allows one to guess the identity of the direction $\underline{n}$
with greater average fidelity than is possible with the states
from $\cal E$. This provides another manifestation of the idea
that the distinguishability in an ensemble can vary while keeping
all pairwise overlaps fixed.

\section{Discussion}

One of the initial motivations for this work was the problem of
determining the optimal compression of quantum information in
mixed state signals \cite{notes,MH1}. For pure states the optimal
compression is given by the von Neumann entropy of the source (and
Schumacher compression \cite{S95,JS} provides an explicit
asymptotically optimal compression protocol) but for mixed states
the value is unknown. One method of compressing the ensemble $\{
\rho_1 , \ldots ,\rho_n ; p_1 , \ldots ,p_n \}$ of mixed states
(where the compresser knows the identity of the signals) is to
first prepare purifications $\ket{\psi_i}$ of the states $\rho_i$
and then apply Schumacher compression to the pure state ensemble
$\{ \ket{\psi_1} , \ldots ,\ket{\psi_n} ; p_1 , \ldots ,p_n \}$.
Thus we wish to construct the ensemble of purifications that has
the least von Neumann entropy. For $n=2$ the solution is given by
the purifications with the largest overlap and the corresponding
minimum entropy can be readily calculated from Uhlmann's
transition probability formula \cite{AU2,JOZ}. However for three
or more states the results in this paper show that the maximum
overlap condition is no longer correct and the problem of
minimising the entropy is evidently more subtle. It has been shown
in \cite{MH2} that the problem of optimal mixed state compression,
in full generality, may be translated into a problem of minimising
the entropy of a suitable ensemble of purifications (of blocks of
signal states).

From a mathematical viewpoint our results amount to an
investigation of the von Neumann entropy $S$ not as a function of
a density matrix, but as a function of variables defining an
ensemble:
\[ S=S(\rho )= S( \sum p_i \proj{\psi_i} ) = S(
 \ket{\psi_1} , \ldots ,\ket{\psi_n} ; p_1 , \ldots ,p_n ). \]
In particular we have considered the behaviour of $S$ when the
states $\ket{\psi_i}$ are varied while the probabilities $p_i$ are
held fixed. In this context many further interesting questions
arise. For example, given an ensemble, what is the largest entropy
that can be attained by deformations that make the states more
parallel? Conversely given a value $S_0$ of von Neumann entropy,
what is the maximum possible (average) overlap of any ensemble
that has entropy $S_0$? Are there ensembles with the property that
every continuous deformation which increases overlaps, begins to
increase the entropy? We may also consider varying the
probabilities. Suppose we have two ensembles ${\cal E} = \{
\ket{\psi_1} , \ket{\psi_2} ,\ket{\psi_3} ; p_1 ,p_2  ,p_3 \}$ and
${\cal E}^* = \{ \ket{\phi_1} , \ket{\phi_2} ,\ket{\phi_3} ; q_1
,q_2  ,q_3 \}$ where the states of ${\cal E}^*$ have greater
pairwise overlap than the corresponding states of $\cal E$. Let
cap($\cal E$) be the maximum entropy attainable from the states in
$\cal E$ by varying the prior probabilities. Is it possible to
have cap$( {\cal E}^*) > {\rm cap}({\cal E})$? This question is of
particular interest since cap($\cal E$) is the classical
information capacity of a quantum channel using the states of
$\cal E$ as basic signal states \cite{HJSWW}.

In the introduction we argued that von Neumann entropy quantifies
a notion of distinguishability for the constituent states of an
ensemble and pointed out that it is surprising that
distinguishability (as well as the lower limit for
compressibility) may be increased by increasing all pairwise
overlaps (for ensembles of three or more states). Thus the notion
of distinguishability and the redundancy involved in quantum
information compression depend not only on the overlaps but also
on the relative phases of the amplitudes $\amp{\psi_i}{\psi_j}$.
For ensembles with just two states there is only one such phase
and it is rendered physically irrelevant by the overall phase
freedom in a physical quantum state. For three or more states the
relative phases cannot be eliminated and they provide further
parameters for issues of compressibility and distinguishability.
For general ensembles we can consider the quantities:
\begin{equation} \label{psik} \Upsilon (i_1 , i_2 , \ldots ,i_k )
= \amp{\psi_{i_1}}{\psi_{i_2}}\amp{\psi_{i_2}}{\psi_{i_3}}\ldots
\amp{\psi_{i_{k-1}}}{\psi_{i_k}}\amp{\psi_{i_k}}{\psi_{i_1}}
\end{equation}
associated to each subset $[\psi_{i_1}], \ldots ,[\psi_{i_k}]$ of
physical states (noting the cyclic chain of indices $i_1 , \ldots
,i_k , i_1$ returning to $i_1$ to complete the cycle.) These
quantities are all unitary invariants and independent of the
choice of phase representatives. Chains of length 1 define the
normalisation while chains of length 2 are just (the squares of)
the pairwise overlaps. The modulus of any chain of length $k$ is a
product of overlaps but its phase is a new quantity. For example
for three states, the phase of any chain of length 3 is (up to a
sign) the parameter $\xi$ considered previously. It would be
interesting to understand the physical bearing of these parameters
on issues of distinguishability and compressibility.

We have seen that a family of ensembles with fixed overlaps can
exhibit a variation of information content. In particular for
ensembles of three states we have the extra quantum mechanical
phase parameter $\xi$. It is interesting to note that an analogous
phenomenon may occur in a purely classical context \cite{Shor}.
Suppose Alice wishes to communicate information to Bob using three
signals A,B,C with equal prior probabilities. The signals
themselves are probability distributions on three values $\{ 1,2,3
\}$. A is the uniform distribution on $\{ 1,2 \}$, B on $\{ 2,3
\}$ and C on $\{ 1,3 \}$. To send a signal, Alice samples the
corresponding distribution and sends the result e.g. to send B she
tosses a fair coin labelled by 2 and 3, and sends the outcome.
(Alternatively we may attribute the probabilistic nature of the
signals to noise in the channel). Bob then reads the received
value. We may readily calculate the probability $p(y|X)$ that Bob
reads $y$ (1,2 or 3) given that Alice sent $X$ (A,B or C) and the
mutual information $I(X:Y)$ between Alice and Bob. In classical
information theory, $I(X:Y)$ quantifies the amount of information
that Bob gets about Alice's message i.e. the information capacity
of the communication protocol. Consider now the same scenario with
three new signals A',B' and C' which are probability distributions
on four values $\{1,2,3,4 \}$. A' is the uniform distribution on
$\{ 1,4 \}$, B' on $\{ 2,4 \}$ and C' on $\{3,4 \}$. The mutual
information is now different but in any reasonable sense, the
signals A', B' and C' have the same pairwise overlaps as the
corresponding signals from the original set A,B,C -- in every case
the two distributions coincide on one value and are disjoint on
the other value. Thus in this purely classical context we again
have the phenomenon that the global information content (in the
sense of mutual information here) differs even though the pairwise
overlaps (and pairwise information contents) are the same.

There are various other possible natural concepts of
distinguishability which one can associate to an ensemble $\{
\ket{\psi_1}, \ldots , \ket{\psi_n}; p_1 , \ldots ,p_n \}$ of pure
quantum states. Two such concepts are the accessible information
and the minimum error probability \cite{helstrom}. For the latter,
we attempt to identify the state by applying a measurement $\cal
M$ with outcomes $1,2, \ldots ,n$. Let $p(j|i)$ be the probability
of obtaining outcome $j$ for the input state $\ket{\psi_i}$. The
error probability is defined by $P_{\rm error} = \sum_i p_i
(1-p(i|i))$. We choose $\cal M$ to minimise $P_{\rm error}$ and
use that minimum value as a measure of distinguishability for the
ensemble (where a smaller value indicates greater
distinguishability). The accessible information of an ensemble is
the maximum Shannon mutual information of any POVM measurement on
the ensemble states. For $n=2$ both of these measures are
monotonic functions of the overlap $|\amp{\psi_1}{\psi_2}|$ but
for $n\geq 3$, like the von Neumann entropy, they are not
functions of the overlaps alone. Thus it seems plausible that they
too -- like the von Neumann entropy -- will exhibit increased
distinguishability with suitable deformations of the ensemble that
decreases all pairwise overlaps. But unlike the von Neumann
entropy this is difficult to study analytically: the computation
of minimum error probability or accessible information involves
difficult optimisations and is generally intractible analytically
for all but the simplest ensembles.

{\bf Acknowledgements:} This work was initiated at the
Elsag-Bailey ISI workshop on quantum computation held in Torino in
1997 and we are grateful for the support and opportunity of
collaboration provided by that meeting. We are especially grateful
to Armin Uhlmann for originally suggesting the Gram matrix
approach adopted in this paper and for other helpful comments.
Much of the work was supported by the European TMR Research
Network ERB-FMRX-CT96-0087. RJ is also supported by the UK
Engineering and Physical Sciences Research Council.

\section{Appendix}
\subsection{Proof of Lemma 1}

Suppose that the two ensembles are unitarily related. Then it is
easy to see that they have the same Gram matrices (as the inner
products are unitary invariants).\\Conversely suppose that
$\mat{G}_1 = \mat{G}_2$. Then the number of states and
corresponding probabilities (being the diagonal of the Gram
matrix) must be the same. So write
\[  {\cal E}_1 = \{
\ket{\alpha_1}, \ldots ,\ket{\alpha_k}; p_1 , \ldots ,p_k \} \] and
\[ {\cal E}_2 = \{ \ket{\beta_1}, \ldots ,\ket{\beta_k}; p_1 ,
\ldots ,p_k \} \]
We will use proof by induction on $k$ to show that $\ket{\alpha_i}
=U\ket{\beta_i}$ for some unitary $U$. The result is clearly true
for $k=1$ i.e. ensembles with only one state. Assume (the induction
hypothesis) that it is true for all ensembles of $k$ states.
Consider two ensembles of $k+1$ states:
\[  {\cal E}_1 = \{
\ket{\alpha_1}, \ldots ,\ket{\alpha_k}, \ket{\alpha_{k+1}};
 p_1 , \ldots ,p_k, p_{k+1} \} \]
\[ {\cal E}_2 = \{ \ket{\beta_1}, \ldots ,\ket{\beta_k},
\ket{\beta_{k+1}}; p_1 , \ldots ,p_k , p_{k+1} \} \]
with the same Gram matrices. Then the sub-ensembles of just the
first $k$ states (with probabilities rescaled by $1/(1-p_{k+1})$)
will have the same Gram matrices so by the induction hypothesis
there is a unitary $U$ with
\[ \ket{\beta_i} = U\ket{\alpha_i} \hspace{1cm} \mbox{for $i=1,
\ldots ,k$ } \]
Now compare the ensembles $U({\cal E}_1 )$ and ${\cal E}_2$. They
differ only in their $(k+1)^{\rm th}$ states which are respectively
$U\ket{\alpha_{k+1}}$ and $\ket{\beta_{k+1}}$. Let $B =
\mbox{span$(\ket{\beta_1} , \ldots , \ket{\beta_k} )$} $. Consider the parallel
and perpendicular projections with respect to $B$:
\[ U\ket{\alpha_{k+1}} = (U\alpha)_\parallel + (U\alpha)_\perp \]
\[ \ket{\beta_{k+1}} = (\beta)_\parallel + (\beta)_\perp \]
Since ${\cal E}_1$ and ${\cal E}_2$ (and hence also $U({\cal E}_1
)$) have equal Gram matrices, $U\ket{\alpha_{k+1}}$ and
$\ket{\beta_{k+1}}$ have equal inner products with $\ket{\beta_1},
\ldots ,\ket{\beta_k}$ and hence with a basis of $B$. Thus the
parallel projections $(U\alpha)_\parallel$ and $(\beta)_\parallel$
are equal and so the perpendicular projections have the same
length. If this length is zero then ${\cal E}_2 = U{\cal E}_1$. If
it is not zero, let $B^\perp$ be the two dimensional space spanned
by $(U\alpha)_\perp$ and $(\beta)_\perp$ and let $V$ be a unitary
transformation in $B^\perp$ with $V(U\alpha)_\perp =
(\beta)_\perp$. Finally let $U'$ be the unitary transformation
which is the identity in $B$ and $V$ in $B^\perp$. Then ${\cal E}_2
= U'U{\cal E}_1$, completing the proof of the lemma.$\Box$

\subsection{Proof of Theorem 1}

For any set of states $\{ [\psi_1 ], [\psi_2 ], [\psi_3 ] \}$ we
can always choose phase representatives making
$\amp{\psi_1}{\psi_2}$ and $\amp{\psi_1}{\psi_3}$ real
non-negative. Hence: \\ $\{ [\psi_1 ], [\psi_2 ], [\psi_3 ] \}$
will have overlaps $a_{12},a_{23},a_{31}$
\\{\bf iff} there are representatives with \begin{equation}
\label{int} \amp{\psi_1}{\psi_2}=a_{12} \hspace{5mm}
\amp{\psi_1}{\psi_3}=a_{31} \hspace{5mm} \mbox{and
$\amp{\psi_2}{\psi_3} = a_{23}e^{i\xi}$ for some $\xi$}
\end{equation} {\bf iff}
\begin{equation}\label{tw} \mat{G}(\xi) \mbox{ defined by }
\frac{1}{3} \left( \begin{array}{ccc} 1 & a_{12} & a_{31} \\
a_{12} & 1 & a_{23}e^{i\xi} \\ a_{31} & a_{23}e^{-i\xi} & 1
\end{array} \right) \mbox{ is a Gram matrix} \end{equation}
{\bf iff} $3\mat{G}$ is a positive matrix (where $\mat{G}$ is the
matrix in eq. (\ref{tw}))
\\ {\bf iff} the eigenvalues $\lambda_1 , \lambda_2 , \lambda_3 $
of $3\mat{G}$ are all non-negative.

Next we claim (and prove in the next paragraph) that the last
condition holds {\bf iff} $\det 3\mat{G} \geq 0$. Then computing
$\det 3\mat{G}$ directly from $\mat{G}$ above, gives the required
condition (\ref{xi}). Note also from (\ref{int}) that $\xi$ is the
phase of the triple quantity $\Upsilon_{123}$ and this completes
the proof of (a).

To justify the claim above, we show that $3\mat{G}$ cannot have
exactly two negative eigenvalues. Consider $\beta = \lambda_1
\lambda_2 + \lambda_2 \lambda_3 + \lambda_3 \lambda_1$ and suppose
that there are two negative eigenvalues $\lambda_1$ and
$\lambda_2$. Then $\tr 3\mat{G} = 3 = \lambda_1 + \lambda_2 +
\lambda_3$ gives \[ \beta = \lambda_1 \lambda_2 +(\lambda_1 +\lambda_2 )
(3 - (\lambda_1 + \lambda_2 )) = -\lambda_1^2 - \lambda_2^2
-\lambda_1 \lambda_2 +3(\lambda_1 +\lambda_2 ) \]
Thus $\beta <0$ as all terms are negative. But $\beta $ is the
linear coefficient in the characteristic equation of $3\mat{G}$ and
a direct calculation of $\det ( \lambda I - 3\mat{G})$ gives
\[ \beta = 3-a_{12}^2 -a_{23}^2 - a_{31}^2 \]
so that $\beta \geq 0$. Hence $3\mat{G}$ can never have exactly two
negative eigenvalues and the non-negativity of $\det 3\mat{G} =
\lambda_1 \lambda_2 \lambda_3 $ is equivalent to the non-negativity
of all three eigenvalues.

To prove (b) we recall that $\Upsilon_{123}$ and hence $\xi$, is an
invariant of any overall unitary transformation. Hence for fixed
$a_{12}, a_{23}$ and $a_{31}$, different $\xi$'s correspond to
unitarily inequivalent sets of states.

To prove (c) suppose that $\{ [\psi_1 ], [\psi_2 ], [\psi_3 ] \}$
is any set of states with the given overlaps. We choose an
orthonormal basis $\{ \ket{0}, \ket{1}, \ket{2} \}$ of ${\cal H}_3$
and phase representatives $\ket{\psi_i}$ as follows. Choose an
arbitrary phase representative $\ket{\psi_1}$ of $[\psi_1 ]$ and
set
\[ \ket{0} = \ket{\psi_1} \]
Choose $\ket{1}$ orthogonal to $\ket{0}$ in the plane of $[\psi_1
]$ and $[\psi_2 ]$ (with phase of $\ket{1}$ to be fixed later).
Then any representative $\ket{\psi_2}$ of $[\psi_2 ]$ has the form
\[ \ket{\psi_2} = a_{12} e^{i\alpha} \ket{0} + \beta \ket{1} \]
Choose the overall phase of $\ket{\psi_2}$ to make $\alpha =0$ and
choose the phase of $\ket{1}$ to make $\beta$ real and
non-negative, so
\[ \ket{\psi_2} = a_{12} \ket{0} + \sqrt{1-a_{12}^2} \ket{1} \]
Then choose $\ket{2}$ orthonormal to $\ket{0}$ and $\ket{1}$ (with
phase to be fixed later) so any representative of $[\psi_3 ]$ has
the form
\[ \ket{\psi_3} = \omega \ket{0} + \eta \ket{1} + \zeta \ket{2} \]
Choose the overall phase of $\ket{\psi_3}$ to make $\omega $ real
non-negative so $\omega = a_{31}$. Choose the phase of $\ket{2}$ to
make $\zeta$ real non-negative. Then
\begin{equation} \label{ppsi3}
\ket{\psi_3} = a_{31} \ket{0} + \eta \ket{1} + z \ket{2}
\hspace{5mm} \mbox{with $z$ real $\geq 0$, $\eta$ complex}
\end{equation}
We have two further conditions:
\begin{equation} \label{a23}
|\amp{\psi_2}{\psi_3}| = a_{23} = |a_{31}a_{12} + \sqrt{1-a_{12}^2}
\eta | \end{equation}
and normalisation:
\begin{equation} \label{norm}
 z^2 = 1-|\eta |^2 -a_{31}^2
\end{equation}  From eq. (\ref{a23}) we introduce $\xi$ so that
\[ a_{31}a_{12} + \sqrt{1-a_{12}^2} \eta = a_{23}e^{i\xi} \]
This gives $\eta$ parameterised by $\xi$:
\begin{equation} \label{eta}
\eta = \frac{ a_{23}e^{i\xi} - a_{31}a_{12}}{\sqrt{1-a_{12}^2}}
\end{equation}
Then eq. (\ref{norm}) gives
\begin{equation} \label{zed}
z^2 = \frac{1-a_{12}^2 - a_{23}^2 - a_{31}^2 +
2a_{12}a_{23}a_{31}\cos \xi }{1-a_{12}^2} =
\frac{ {\rm det} \left( \begin{array}{ccc} 1 & a_{12} & a_{31} \\
a_{12} & 1 & e^{i\xi}a_{23} \\ a_{31} & e^{-i\xi}a_{23} & 1
\end{array} \right) }{1-a_{12}^2} \end{equation}
Substituting eq. (\ref{eta}) and eq. (\ref{zed}) into eq.
(\ref{ppsi3}) gives the required form of $\ket{\psi_3}$. Note that
the condition $z^2 \geq 0$ in eq. (\ref{zed}) reproduces the
condition (\ref{xi}) in part (a) of the theorem.$\Box$

\subsection{Proof of Theorem 2}

Given the ensemble ${\cal E}=
 \{ \ket{\psi_1}
,\ket{\psi_2}, \ket{\psi_3}; p_1 ,p_2 ,p_3 \}$ let ${\cal E}(\xi
)$ denote the family of ensembles (up to unitary equivalence)
which have the same overlaps as $\cal E$. Here $\xi$ is the phase
of $\Upsilon_{123}$. We study the variation of $S$ with $\xi$
through a sequence of lemmas (with proofs given at the end):

{\bf Lemma A1:} The ensembles ${\cal E}(\xi )$ have constant $\tr
\rho$ and constant $\tr \rho^2$. Hence $S$ may be viewed as a
function of $\tr \rho^3 = \tr \mat{G}^3$.

{\bf Lemma A2:} If $\tr \rho$ and $\tr \rho^2$ are held constant
then $S$ is a monotonically increasing function of $\tr \rho^3$.

{\bf Lemma A3:} For any ensemble $\cal E$, $\tr \rho^3$ has the
form
\begin{equation} \label{tr3}
\tr \rho^3 = C+ 6p_1 p_2 p_3 {\rm Re}\Upsilon_{123} = C+6 p_1 p_2
p_3 a_{12}a_{23} a_{31} \cos \xi \end{equation} where $C$ is
independent of $\xi$.

In view of these lemmas we have:

{\bf Proof of theorem 2:} We consider the family of ensembles
${\cal E}(\xi)$ with the same overlaps as $\cal E$, parameterised
by $\xi$, the phase of $\Upsilon_{123}$.
\\(a) If $\Upsilon$ is not real and $\xi_0 \neq \pm
\xi_{max}$ (as $\cal E$ has rank 3) then we can perturb $\xi_0$
slightly in both the positive and negative directions inducing
either an increase or decrease in $\tr \rho^3$ (by lemma A3).
Hence by lemmas A1 and A2 we may either increase or decrease $S$
while keeping the overlaps the same, giving the possibilities {\bf
(D1)} and {\bf (D2)}.\\(b) If $\xi_0 =0$ then $\cos \xi_0$ is at
its maximum value. Thus any perturbation of $\xi_0$ leads to a
decrease in $\tr \rho^3$ and hence a decrease in $S$ (while
keeping all overlaps constant). Thus we get {\bf (D2)} only. At
$\xi_0 = \pi$, $\cos \xi_0$ is minimum so similarly, we get only
{\bf (D1)}.$\Box$

{\bf Proof of Lemma A1:} We have $\tr \rho =1$ for all $\xi$. Also
$\tr \rho^2 = \tr \mat{G}^2 = \sum_{ij} G_{ij}G_{ji} = \sum
|G_{ij}|^2$ as $\mat{G}$ is Hermitian. Since ${\cal E}(\xi)$ have
constant overlaps, the Gram matrices are related by Hadamard
product with $\mat{r}$ of the form in eq. (\ref{rphi}) with
$|r_{ij}|=1$ for all $i,j$. Hence $\tr \rho^2$ remains constant.
The three eigenvalues of $\rho$ are uniquely determined by the
values of $\tr \rho$, $\tr \rho^2$ and $\tr \rho^3$ so $S$ may be
viewed as a function of $\tr \rho^3$. $\Box$

{\bf Proof of Lemma A2:} Note first that $S$ is unitarily
invariant so we may assume without loss of generality that $\rho$
is diagonal, and work on the classical probability simplex
\[ {\cal P}_3 = \{ (\lambda_1 , \lambda_2 , \lambda_3 ) :
\lambda_1 \geq 0, \lambda_2 \geq 0, \lambda_3 \geq 0 \hspace{3mm}
\mbox{and $\lambda_1 +\lambda_2 +\lambda_3 =1$} \} \] To represent
the constraint $\tr \rho^2 = {\rm const}$ it is convenient to
introduce a polar co-ordinate system $(r,\theta)$ in ${\cal P}_3$
as shown in figure 2. $r$ is measured from the centre $M =
(\frac{1}{3},\frac{1}{3},\frac{1}{3})$ of ${\cal P}_3$ and
$\theta$ is measured anticlockwise from the line joining $M$ to
the vertex (1,0,0).

\begin{figure}
 \centerline{  \psfig{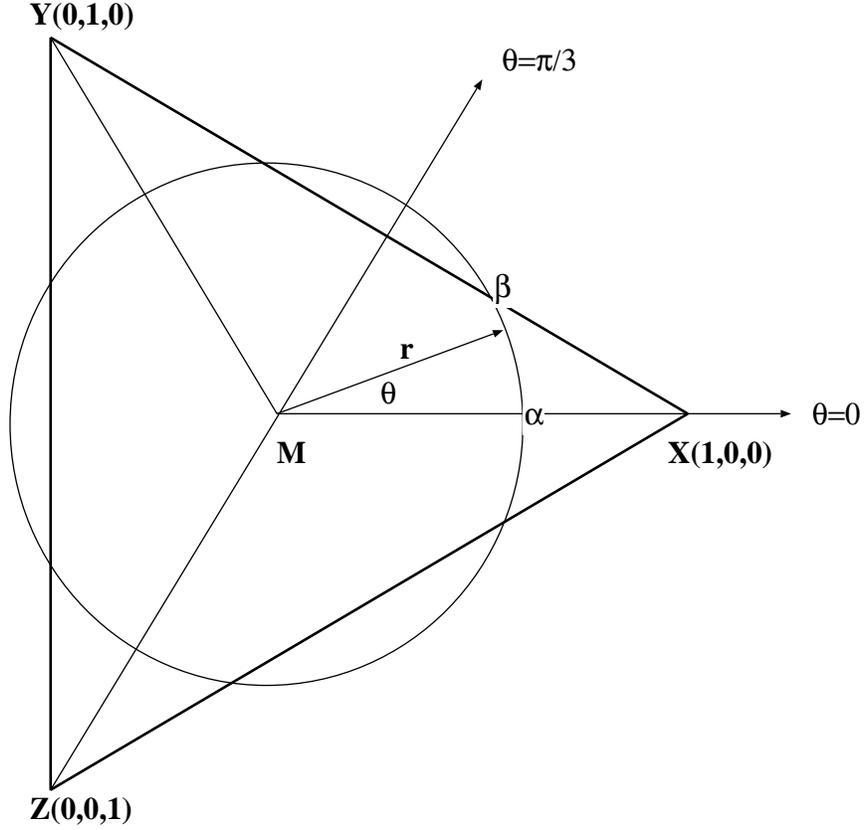}}
\caption{The triangle {\bf XYZ} is the probability simplex ${\cal
P}_3$ (or space of diagonal density matrices diag$(\lambda_1
,\lambda_2 , \lambda_3 )$) in $d=3$ dimensions. It is obtained by
intersecting the plane $\lambda_1 + \lambda_2 + \lambda_3 = 1$
with the positive octant in the space of all triples $(\lambda_1 ,
\lambda_2 ,\lambda_3 )$. The three vertices are the three pure
states and the edges are rank 2 diagonal states. The central point
$M= {\rm diag}( \frac{1}{3},\frac{1}{3},\frac{1}{3})$ is the
maximally mixed state. Along lines $MX$. $MY$ and $MZ$ two
eigenvalues are equal. We introduce polar co-ordinates on this
simplex with $r$ being the distance from $M$ and the polar angle
$\theta$ is measured anticlockwise from the line $MX$. The
co-ordinates for a general state are given in eq. (\ref{lam}). The
arc $\alpha\beta$ includes all diagonal states $\rho$  (up to a
permutation of the diagonal entries) at a constant radius $r$ with
a fixed $\tr \rho^2 = \frac{1}{3}+r^2$ (cf eq. (\ref{two}) {\em et
seq.}) }
  \label{bild1}
 \end{figure}
 A direct calculation gives the co-ordinates of a general state as:
\begin{equation} \label{lam}
\begin{array}{l}
\lambda_1 = \frac{1}{3} + \sqrt{\frac{2}{3}} r\cos \theta \\
\lambda_2 = \frac{1}{3} + \sqrt{\frac{2}{3}} r\cos (\theta +
\frac{2\pi}{3}) \\ \lambda_3 = \frac{1}{3} + \sqrt{\frac{2}{3}}
r\cos (\theta + \frac{4\pi}{3})
\end{array}
\end{equation}
The constraint
\begin{equation} \label{one} \tr \rho^2 = \sum \lambda_i^2 = A^2 = {\rm const}
\end{equation}
corresponds to the intersection of the simplex $XYZ$ with a sphere
of radius $A$ centred at (0,0,0). This gives a circle or part of a
circle (as shown in figure 2) within the simplex. In terms of
polar co-ordinates we get
\begin{equation}\label{two} \sum \lambda_i^2 = \frac{1}{3} +r^2 \end{equation}
In figure 2 the lines $MX$, $MY$ and $MZ$ divide the circle into
six symmetrical parts corresponding to the six possible
permutations of the eigenvalues. Since we will be interested only
in the (un-ordered) set of values of the $\lambda$'s it suffices
to consider just one of these regions. Thus without loss of
generality we may take
\[ 0\leq \theta \leq \theta_{max} \leq \frac{\pi}{3} \]
For some values of constant $\tr \rho^2$ or $r$, the angle
$\theta$ will have a maximum value $\theta_{max}$ smaller than
$\pi/3$ e.g. for the $r$ shown in figure 2 we take only the arc
$\alpha\beta$. At the point $\beta$ one of the eigenvalues has
become zero.

From eq. (\ref{two}) we see that the constraint $\tr \rho^2 = {\rm
const}$ corresponds to $r$ being constant. Using eq. (\ref{lam})
we get:
\[ \tr \rho^3 = \sum \lambda_i^3 = \frac{1}{9} + r^2 +
\frac{1}{\sqrt{6}} r^3 \cos 3\theta \] Since $\theta$ lies in the
range $[0, \frac{\pi}{3}]$ we see that $\tr \rho^3$ is a
monotonically decreasing function of $\theta$.

Next we show that $S$ is also monotonically decreasing with
$\theta$. To study the variation of entropy
\[ S = - \sum \lambda_i \log \lambda_i \]
with $\theta$ when $\tr \rho^2 $ is held constant, note that $\sum
\lambda_i = 1$ so we get
\begin{equation} \label{three}
 \sum \frac{\partial \lambda_i}{\partial \theta} = 0 \end{equation}
and so
\[ \frac{\partial S}{\partial \theta} =
- \sum_i \frac{\partial \lambda_i}{\partial \theta}\log \lambda_i
\] Since $0\leq \theta\leq \frac{\pi}{3}$ we have from eq.
(\ref{lam})
\[ \frac{\partial\lambda_1}{\partial\theta}\, , \,
\frac{\partial \lambda_2}{\partial\theta} \leq 0 \hspace{1cm}
\frac{\partial \lambda_3}{\partial\theta} \geq 0 \] Hence using
eq. (\ref{three}) we get
\[ \frac{\partial S}{\partial\theta} = | \frac{\partial \lambda_1}{\partial
\theta}| \log \frac{\lambda_1}{\lambda_3} + | \frac{\partial
\lambda_2}{\partial\theta}| \log \frac{\lambda_2}{\lambda_3} \]
Using the inequality $\log x \leq x-1$ we get directly:
\[ \frac{\partial S}{\partial\theta} \leq 0 \]
with equality possible only for $\theta = 0$ or $\pi/3$.

To summarise, any given $\rho$ corresponds to a unique point
$(r,\theta)$ with $0\leq \theta\leq \pi/3$. If $\tr \rho^2 =
\frac{1}{3}+r^2$ is held constant then both $\tr \rho^3$ and $S$
are monotonically decreasing functions of $\theta$. Hence $S$ is a
monotonically increasing function of $\tr \rho^3$ which completes
the proof of Lemma A2. $\Box$

{\bf Proof of Lemma A3:} We have
\begin{equation} \label{trg3}
\tr \rho^3 = \tr \mat{G}^3 = \sum_{ijk} G_{ij}G_{jk}G_{ki}
\end{equation}
Since $\rho$ is invariant under choices of phase representatives
we may assume without loss of generality that $\mat{G}$ has the
form given in eq. (\ref{gramform}). Then in eq. (\ref{trg3}) the
$\xi$ dependence arises only through three terms (via cyclic
permutation of the subscripts) of the form $G_{23}G_{31}G_{12} =
p_1 p_2 p_3 \Upsilon_{123}$ and the three corresponding complex
conjugate terms $G_{32}G_{21}G_{13}= p_1 p_2 p_3
\overline{\Upsilon_{123}}$. This gives eq. (\ref{tr3}).$\Box$


\begin{thebibliography}{xx}
\bibitem{peres} A. Peres, Phys. Lett. A {\bf 128}, 19 (1988).
\bibitem{Cover} T. M. Cover and J. A. Thomas, {\it Elements of Information
Theory} (John Wiley and Sons, N.Y. 1991).
\bibitem{S95}
B. Schumacher, Phys. Rev. A {\bf 51}, 2738 (1995).
\bibitem{JS}
R. Jozsa and B. Schumacher, J. Mod. Optics {\bf 41}, 2343 (1994).
\bibitem{HJSWW} P. Hausladen, R. Jozsa, B. Schumacher, M.
Westmoreland and W. Wootters, Phys. Rev. A {\bf 54}, 1869 (1996).
\bibitem{Barnum}
H. Barnum, Ch. Fuchs, R. Jozsa and B. Schumacher, Phys. Rev. A {\bf 54}, 4707
(1996).
\bibitem{HJW} L. Hughston, R. Jozsa and W. Wootters, Phys. Lett. A
{\bf 183}, 14 (1993).
\bibitem{Wehrl} A. Wehrl, Rev. Mod. Phys. {\bf 50}, 221 (1978.
\bibitem{HJ}R. Horn and C. Johnson, {\it Matrix Analysis}
(Cambridge University Press 1985).
\bibitem{uhl77} A. Uhlmann, Comm. Math. Phys. {\bf 54}, 21 (1977).
\bibitem{lindblad} G. Lindblad, Commun. Math. Phys. {\bf 40}, 147
(1975).
\bibitem{notes} H. Barnum, C. Caves, Ch. Fuchs, R. Jozsa and B.
Schumacher, "Quantum Coding for Mixed States" (unpublished
amnuscript, September 1995).
\bibitem{MH1} M. Horodecki, Phys. Rev. A {\bf 57}, 3363 (1998).
\bibitem{AU2} A. Uhlmann, Rep. Math. Phys. {\bf 9}, 273 (1976).
\bibitem{JOZ} R. Jozsa, J. Mod. Optics {\bf 41}, 2315 (1994).
\bibitem{MH2} M. Horodecki, "Towards optimal compression for mixed
signal states", preprint available at
http://xxx.lanl.gov/quant-ph/9905058 (1999).
\bibitem{MP} N. Gisin and S. Popescu, "Spin flips and quantum
information for anti-parallel spins" preprint available at
http://xxx.lanl.gov/quant-ph/9901072 (1999).
\bibitem{helstrom} C. Helstrom, {\it Quantum Detection and
Estimation Theory} (Academic Press New York, 1976), chapter 4.
\bibitem{Shor} Peter Shor, private communication.
\end{thebibliography}
\end{document}